\documentclass[prd,aps,showpacs,nofootinbib,preprint,eqsecnum]{revtex4}
\usepackage{graphicx,color,amsmath,amsxtra}
\usepackage{epsf}
\usepackage{amssymb}
\usepackage{enumerate}
\usepackage{hhline}
\usepackage{array}
\usepackage{tabularx}

\newcommand{\bear}{\begin{array}}  \newcommand{\eear}{\end{array}}
\newcommand{\bea}{\begin{eqnarray}}  \newcommand{\eea}{\end{eqnarray}}
\newcommand{\beq}{\begin{equation}}  \newcommand{\eeq}{\end{equation}}
\newcommand{\bef}{\begin{figure}}  \newcommand{\eef}{\end{figure}}
\newcommand{\bec}{\begin{center}}  \newcommand{\eec}{\end{center}}

\newcommand{\Eqn}[1]{&\hspace{-0.2em}#1\hspace{-0.2em}&}

\def\Vec#1{\mbox{\boldmath $#1$}}

\def\be{\begin{equation}}
\def\ee{\end{equation}}
\def\bea{\begin{eqnarray}}
\def\eea{\end{eqnarray}}
\def\beq{\begin{eqnarray}}
\def\eeq{\end{eqnarray}}

\def\be{\begin{equation}}
\def\ee{\end{equation}}
\def\bea{\begin{eqnarray}}
\def\eea{\end{eqnarray}}
\def\beq{\begin{eqnarray}}
\def\eeq{\end{eqnarray}}

\begin{document}

\title{Behavior of $F(R)$ gravity around a crossing of the phantom divide
}

\author{Kazuharu Bamba\footnote{E-mail address: bamba@phys.nthu.edu.tw} 
}
\affiliation{
Department of Physics, National Tsing Hua University, Hsinchu, Taiwan 300
}


\begin{abstract} 

We study a model of $F(R)$ gravity in which a crossing of the phantom 
divide can be realized. 
In particular, we demonstrate the behavior of $F(R)$ gravity around a 
crossing of the phantom divide by taking into account the presence of cold 
dark matter. 

\end{abstract}


\maketitle

\section{Introduction}

It is observationally supported that the current expansion of the universe 
is accelerating~\cite{WMAP, SN1}. 
The scenarios to account for the current accelerated expansion of the universe 
fall into two broad categories~\cite{Peebles, Copeland:2006wr, 
DM, Nojiri:2006ri, rv-2, Sotiriou:2008rp, Caldwell:2009ix, 
Silvestri:2009hh, Sami:2009jx}. 
One is to introduce ``dark energy'' in the framework of general 
relativity. The other is to study a modified gravitational theory, e.g., 
$F(R)$ gravity, in which the action is represented by an 
arbitrary function $F(R)$ of the scalar curvature $R$ (for reviews, 
see~\cite{Nojiri:2006ri, rv-2, Sotiriou:2008rp, Caldwell:2009ix, 
Silvestri:2009hh, Sami:2009jx}; 
and for a new approach, see~\cite{Chakrabarti:2009ku}). 

On the other hand, various observational data~\cite{observational status} 
imply that the ratio of the effective pressure to the effective energy density 
of the universe, i.e., the effective equation of state (EoS) 
$w_\mathrm{eff}\equiv p_\mathrm{eff}/\rho_\mathrm{eff}$, 
may evolve from larger than $-1$ (non-phantom phase) to less 
than $-1$ (phantom phase~\cite{phantom, Cline:2003gs}). Namely, 
it crosses $-1$ (the phantom divide) at the present time or in the near past. 
A number of models to realize the crossing of the phantom divide have 
been proposed (for a detailed review, see~\cite{Copeland:2006wr}). 
 
There are also several studies for the crossing of the phantom divide 
in the framework of $F(R)$ gravity~\cite{Nojiri:2006ri, Abdalla:2004sw, 
Amendola:2007nt, Brevik, Elizalde:2009gx, Sadeghi:2009qr, Linder:2009jz, 
Setare:2009jf}\footnote{The equivalence between $F(R)$ 
gravity and the scalar-tensor theory has been indicated in Ref.~\cite{EMS}. 
The crossing of the phantom divide in scalar-tensor theories has been 
investigated in Ref.~\cite{scalar-tensor theories}.} 
(for related works, see~\cite{B-Wang}). 
An explicit model with realizing a crossing of the phantom divide 
has been constructed in Ref.~\cite{Bamba:2008hq} 
and its thermodynamics has been examined~\cite{Bamba:2009ay}. 
Moreover, in Ref.~\cite{Bamba:2009kc} it has been illustrated that multiple 
crossings of the phantom divide can occur in $F(R)$ gravity as 
the scalar field theories such as an oscillating quintom 
model~\cite{Feng:2004ff} or a quintom with two scalar 
fields~\cite{Zhang:2005eg} in the framework of general relativity 
(see also~\cite{M-Li, Cai:2009zp}). 

In this paper, we study a model of $F(R)$ gravity in which 
a crossing of the phantom divide can be realized by taking into account the 
presence of cold dark matter. We demonstrate the behavior of $F(R)$ gravity 
around a crossing of the phantom divide. 
In our previous work~\cite{Bamba:2008hq}, 
an analytic solution of $F(R)$ gravity to realize a crossing of the 
phantom divide without matter has been derived. 
In this work, as a further investigation, we examine a solution of $F(R)$ 
gravity to realize a crossing of the phantom divide with cold dark matter. 
We use units of $k_\mathrm{B} = c = \hbar = 1$ and denote the
gravitational constant $8 \pi G$ by 
${\kappa}^2 \equiv 8\pi/{M_{\mathrm{Pl}}}^2$ 
with the Planck mass of $M_{\mathrm{Pl}} = G^{-1/2} = 1.2 \times 10^{19}$GeV.

The paper is organized as follows. 
In Sec.\ II, we explain the reconstruction method of $F(R)$ gravity 
proposed in Ref.~\cite{RM}. 
In Sec.\ III, we study a model of $F(R)$ with realizing a crossing of 
the phantom divide by using the reconstruction method and 
investigate the behavior of $F(R)$ gravity around 
a crossing of the phantom divide. 
Finally, conclusions are given in Sec.\ IV.

\section{Reconstruction method}

First, we explain the reconstruction method of $F(R)$ 
gravity proposed in Ref.~\cite{RM}. 
The action of $F(R)$ gravity with matter is as follows: 
\begin{eqnarray}
S = \int d^4 x \sqrt{-g} \left[ \frac{F(R)}{2\kappa^2} +
{\mathcal{L}}_{\mathrm{matter}} \right]\,,
\label{eq:2.1}
\end{eqnarray}
where $g$ is the determinant of the metric tensor $g_{\mu\nu}$ and
${\mathcal{L}}_{\mathrm{matter}}$ is the matter Lagrangian. 
By using proper functions $P(\phi)$ and $Q(\phi)$ of a scalar field $\phi$, 
the action in Eq.~(\ref{eq:2.1}) can be rewritten to 
\begin{eqnarray}
S=\int d^4 x \sqrt{-g} \left\{ \frac{1}{2\kappa^2} \left[ P(\phi) R + Q(\phi)
\right] + {\mathcal{L}}_{\mathrm{matter}} \right\}\,.
\label{eq:2.2}
\end{eqnarray}
The scalar field $\phi$ may be regarded as an auxiliary scalar field because 
it has no kinetic term. From Eq.~(\ref{eq:2.1}), the equation of 
motion of $\phi$ is given by 
\begin{eqnarray}
0=\frac{d P(\phi)}{d \phi} R + \frac{d Q(\phi)}{d \phi}\,.
\label{eq:2.3}
\end{eqnarray}
Substituting $\phi=\phi(R)$ into the action in Eq.~(\ref{eq:2.2}) yields 
the expression of $F(R)$ as 
\begin{eqnarray}
F(R) = P(\phi(R)) R + Q(\phi(R))\,.
\label{eq:2.4}
\end{eqnarray}
 From Eq.~(\ref{eq:2.2}), the field equation of modified gravity is 
derived as
\begin{eqnarray}
\frac{1}{2}g_{\mu \nu} \left[ P(\phi) R + Q(\phi) \right]
-R_{\mu \nu} P(\phi) -g_{\mu \nu} \Box P(\phi) +
{\nabla}_{\mu} {\nabla}_{\nu}P(\phi) + \kappa^2
T^{(\mathrm{matter})}_{\mu \nu} = 0\,,
\label{eq:2.5}
\end{eqnarray}
where ${\nabla}_{\mu}$ is the covariant derivative operator associated with 
$g_{\mu \nu}$, $\Box \equiv g^{\mu \nu} {\nabla}_{\mu} {\nabla}_{\nu}$ 
is the covariant d'Alembertian for a scalar field, and 
$T^{(\mathrm{matter})}_{\mu \nu}$ is the contribution to 
the energy-momentum tensor form matter.

We assume the flat 
Friedmann-Robertson-Walker (FRW) space-time with the metric,
\begin{eqnarray}
{ds}^2 = -{dt}^2 + a^2(t)d{\Vec{x}}^2\,,
\label{eq:2.6}
\end{eqnarray}
where $a(t)$ is the scale factor. 
In this background, the components of
$(\mu,\nu)=(0,0)$ and $(\mu,\nu)=(i,j)$ $(i,j=1,\cdots,3)$ 
in Eq.~(\ref{eq:2.5}) read 
\begin{eqnarray}
&&
-6H^2P(\phi(t)) -Q(\phi(t)) -6H \frac{dP(\phi(t))}{dt} + 2\kappa^2\rho = 0\,,
\label{eq:2.7} \\
&&
2\frac{d^2P(\phi(t))}{dt^2}+4H\frac{dP(\phi(t))}{dt}+
\left(4\dot{H}+6H^2 \right)P(\phi(t)) +Q(\phi(t)) + 2\kappa^2 p = 0\,,
\label{eq:2.8}
\end{eqnarray} 
where $H=\dot{a}/a$ is the Hubble parameter with 
$\dot{~}=\partial/\partial t$ and 
$\rho$ and $p$ are the sum of the energy density and
pressure of matters with a constant EoS $w_i$, respectively, 
with $i$ being some component of matters. 
After eliminating $Q(\phi)$ from Eqs.~(\ref{eq:2.7}) and (\ref{eq:2.8}), 
we get 
\begin{eqnarray}
\frac{d^2P(\phi(t))}{dt^2} -H\frac{dP(\phi(t))}{dt} +2\dot{H}P(\phi(t)) +
\kappa^2 \left( \rho + p \right) = 0\,.
\label{eq:2.9}
\end{eqnarray}
The scalar field $\phi$ may be taken as $\phi = t$ 
if it is redefined properly. 
By representing $a(t)$ as
\begin{eqnarray}
a(t) = \bar{a} \exp \left( \tilde{g}(t) \right)
\label{eq:2.10}
\end{eqnarray}
in terms of a constant of $\bar{a}$ and  a proper function of $\tilde{g}(t)$
and using $H= d \tilde{g}(\phi)/\left(d \phi \right)$, we rewrite 
Eq.~(\ref{eq:2.9}) to be 
\begin{eqnarray}
&&
\frac{d^2P(\phi)}{d\phi^2} -\frac{d \tilde{g}(\phi)}{d\phi}
\frac{dP(\phi)}{d\phi} +2 \frac{d^2 \tilde{g}(\phi)}{d \phi^2}
P(\phi) \nonumber \\
&& \hspace{10mm}
{}+
\kappa^2 \sum_i \left( 1+w_i \right) \bar{\rho}_i
\bar{a}^{-3\left( 1+w_i \right)} \exp
\left[ -3\left( 1+w_i \right) \tilde{g}(\phi) \right] = 0\,,
\label{eq:2.11}
\end{eqnarray}
where $\bar{\rho}_i$ is a constant. 
Moreover, from Eq.~(\ref{eq:2.7}), we obtain
\begin{eqnarray}
Q(\phi) \Eqn{=} -6 \left[ \frac{d \tilde{g}(\phi)}{d\phi} \right]^2 P(\phi)
-6\frac{d \tilde{g}(\phi)}{d\phi} \frac{dP(\phi)}{d\phi} \nonumber \\
&& \hspace{10mm}
{}+
2\kappa^2 \sum_i \bar{\rho}_i \bar{a}^{-3\left( 1+w_i \right)}
\exp
\left[ -3\left( 1+w_i \right) \tilde{g}(\phi) \right]\,.
\label{eq:2.12}
\end{eqnarray}

We note that if we redefine the auxiliary scalar field $\phi$
by $\phi=\Phi(\varphi)$ with a proper function $\Phi$ and define 
$\tilde P(\varphi)\equiv P(\Phi(\varphi))$ and
$\tilde Q(\varphi)\equiv Q(\Phi(\varphi))$, the new action 
\begin{eqnarray}
S \Eqn{=} \int d^4 x \sqrt{-g}\left[ \frac{\tilde F(R)}{2\kappa^2} 
+{\cal L}_{\rm matter} \right]\,, 
\label{eq:2.13} \\
\tilde F(R) \Eqn{\equiv} \tilde P(\varphi) R + \tilde Q(\varphi)\,, 
\label{eq:2.14}
\end{eqnarray}
is equivalent to the action in Eq.~(\ref{eq:2.2}) because 
$\tilde F(R) = F(R)$. 
Here, $\varphi$ is the inverse function of $\Phi$ and 
we can solve $\varphi$ with respect to $R$ as 
$\varphi=\varphi(R) = \Phi^{-1}(\phi(R))$ by using $\phi = \phi(R)$. 
Consequently, we have the choices in $\phi$ like a gauge symmetry 
and therefore we can identify $\phi$ with time $t$, i.e., $\phi=t$, which can 
be interpreted as a gauge condition corresponding to the reparameterization of 
$\phi=\phi(\varphi)$~\cite{Bamba:2008hq}. 
Thus, if we have the relation $t = t(R)$, in principle we can obtain the 
form of $F(R)$ by solving Eq.~(\ref{eq:2.11}) with Eqs.~(\ref{eq:2.4}) and 
(\ref{eq:2.12}). 

We also remark that a crossing of the phantom divide 
cannot be described by a naive model of $F(R)$ gravity. 
To realize the crossing, $F(R)$ needs to be a double-valued function, 
where the cut could correspond to $w_\mathrm{eff} = -1$. 
However, the crossing can be performed by the extension of $F(R)$ gravity, 
whose action is given by $P(\phi) R + Q(\phi)$.

\section{Model}

Next, we examine a model of $F(R)$ with realizing a crossing of 
the phantom divide by using the reconstruction method and 
investigate the behavior of $F(R)$ gravity around 
a crossing of the phantom divide.

\subsection{Crossing of the phantom divide}

To illustrate the behavior of $F(R)$ with realizing a crossing of the phantom 
divide, we consider the case in which the Hubble 
rate $H(t)$ is expressed as~\cite{Nojiri:2005sx}
\begin{eqnarray}
H = n\left( \frac{1}{t} + \frac{1}{t_s - t} \right)\,,
\label{eq:3.1} 
\end{eqnarray}
where $n$ is a positive constant and $t_\mathrm{s}$ is the time when 
the Big Rip singularity~\cite{BR} appears as will be shown 
later\footnote{Other kinds of finite-time future singularities have been 
studied in Ref.~\cite{sudden}.}. 

In the FRW background, the effective 
energy density and pressure of the universe are given by 
$\rho_\mathrm{eff} = 3H^2/\kappa^2$ and
$p_\mathrm{eff} = -\left(2\dot{H} + 3H^2 \right)/\kappa^2$, respectively. 
The effective EoS $w_\mathrm{eff} = p_\mathrm{eff}/\rho_\mathrm{eff}$ 
is defined as~\cite{Nojiri:2006ri}
\begin{eqnarray}
w_\mathrm{eff} \equiv -1 -\frac{2\dot{H}}{3H^2}\,, 
\label{eq:3.2}
\end{eqnarray}
which implies that a crossing of the phantom divide 
occurs when the sign of $\dot{H}$ changes. 

When $t \to 0$, i.e., $t \ll t_s$, $H(t)$ in Eq.~(\ref{eq:3.1}) behaves as 
$
H(t) \sim n/t 
$ 
and therefore $\dot{H} \sim -n/t^2 <0$. 
In this limit, it follows from Eq.~(\ref{eq:3.2}) that
the effective EoS is given by
$
w_\mathrm{eff} = -1 + 2/\left(3n\right) > -1.
$ 
Such behavior is identical with that in the Einstein gravity 
with matter whose EoS is greater than $-1$. This is the non-phantom phase. 
On the other hand, when $t\to t_s$, we find
$
H(t) \sim n/\left(t_s - t\right) 
$
and hence $\dot{H} \sim n/\left(t_s - t\right)^2 >0$. 
We only consider the period $0<t<t_s$ because $H$ should be positive. 
In this case, the scale factor is given by 
$a(t) \sim \bar{a} \left( t_s - t \right)^{-n}$. 
Thus, when $t\to t_s$, $a \to \infty$, namely, the Big Rip singularity 
appears. 
In this limit, the effective EoS is given by
$
w_\mathrm{eff} = -1 - 2/\left(3n\right) <-1. 
$ 
Such behavior is identical with the case in which there is a phantom matter
with its EoS being smaller than $-1$. This is the phantom phase. 
Moreover, from Eq.~(\ref{eq:3.2}), we see that the effective EoS 
$w_\mathrm{eff}$ becomes $-1$ when $\dot{H}=0$. 
Solving $w_\mathrm{eff} = -1$ with respect to 
$t$ by using Eq.~(\ref{eq:3.1}), 
we find that the effective EoS crosses the phantom divide at 
$t=t_\mathrm{c}$ given by $t_\mathrm{c} =t_s/2$. 
As a consequence, 
in case of Eq.~(\ref{eq:3.1}), a crossing of the phantom divide 
can occur. 
We show the time evolution of $w_\mathrm{eff}$ in Fig.~1 with 
$\tilde{t} \equiv t/t_0$, where $t_0$ is the present time. 
In all figures, we take $n=10$ and 
$t_s \equiv \alpha t_0$ with $\alpha = 2.0$. 
In this case, $t_\mathrm{c} = t_0$. 
From Fig.~1, we see that 
at the present time, a crossing of the phantom divide 
from the non-phantom phase ($w_\mathrm{eff} > -1$) to the phantom one 
($w_\mathrm{eff} < -1$) can be realized. 

\begin{figure}[tbp]
\begin{center}
   \includegraphics{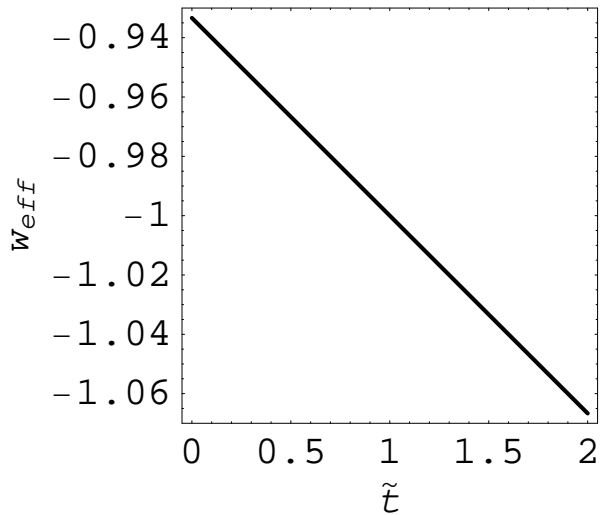}
\caption{Time evolution of $w_\mathrm{eff}$ for $n=10$ and $\alpha = 2.0$ 
with $\tilde{t} = t/t_0$. 
}
\end{center}
\label{fg:1}
\end{figure}

\subsection{Behavior of $F(R)$ gravity around a crossing of the phantom divide}

In what follows, we take $\phi = t$. 
From Eqs.~(\ref{eq:2.10}) and (\ref{eq:3.1}), 
$H= d \tilde{g}(t)/\left(d t \right)$ and $R=6\left( \dot{H} + 2H^2 \right)$, 
we obtain 
\begin{eqnarray} 
\tilde{g}(t) \Eqn{=} n\ln \left( \frac{t}{t_s - t} \right)\,, 
\label{eq:3.3} \\
a(t) \Eqn{=} \left[ \frac{\left(\alpha -1\right)t}{t_s - t} \right]^n\,,
\label{eq:3.4} \\
R \Eqn{=} \frac{6nt_s}{t^2 \left( t_s - t \right)^2} 
\left[ \left( 2n-1 \right)t_s +2t \right]\,,
\label{eq:3.5}
\end{eqnarray}
where we have taken 
$\bar{a}= \left(\alpha -1\right)^n$ so that the present value of 
the scale factor should be unity.

We define $X \equiv t/t_s$ and solve Eq.~(\ref{eq:3.5}) with 
respect to $X$. If $n$ is much larger than unity, we can neglect
the second term on the right-hand side of Eq.~(\ref{eq:3.5}) and therefore 
obtain the approximate solutions 
\begin{eqnarray}
X (\tilde{R}) \approx \frac{1\pm\sqrt{1-4\tilde{R}^{-1/2}}}{2}\,,
\label{eq:3.6}
\end{eqnarray}
where
\begin{eqnarray}
\tilde{R} \Eqn{=} \frac{t_s^2R}{6n\left( 2n-1 \right)}\,.
\label{eq:3.7}
\end{eqnarray} 
For the lower sign and the upper one in Eq.~(\ref{eq:3.6}), 
$X$ varies as $0<X \leq 1/2$ and $1/2 \leq X <1$, respectively. 

For simplicity, we consider the case in which 
there exists a matter with a constant EoS $w=p/\rho$. 
In this case, 
Eqs.~(\ref{eq:2.11}) and (\ref{eq:2.12}) are rewritten to 
\begin{eqnarray}
&&
4\tilde{R}^{5/2} \left( 1-4\tilde{R}^{-1/2} \right) 
\frac{d^2 P(\tilde{R})}{d \tilde{R}^2}
\nonumber \\
&& 
{}+2\tilde{R}^{2} \left[ \left( 3-10\tilde{R}^{-1/2} \right)
\mp n \sqrt{1-4\tilde{R}^{-1/2}} \, \right] 
\frac{d P(\tilde{R})}{d \tilde{R}} \pm 2n\tilde{R} 
\sqrt{1-4\tilde{R}^{-1/2}} P(\tilde{R}) 
\nonumber \\
&& 
{}+ t_s^2 \kappa^2 (1+w) \bar{\rho} 
\left[ \left( \alpha-1 \right) \left( 
\frac{1 \pm \sqrt{1-4\tilde{R}^{-1/2}}}{1 \mp \sqrt{1-4\tilde{R}^{-1/2}}} 
\right) \right]^{-3n(1+w)} 
=0
\label{eq:3.8}
\end{eqnarray}
and
\begin{eqnarray}
\tilde{t}_s^{\,2} Q(\tilde{R})
\Eqn{=} 
-\frac{n}{2n-1} \tilde{R} P(\tilde{R}) 
\mp \frac{2}{2n-1} \tilde{R}^2  \sqrt{1-4\tilde{R}^{-1/2}} 
\frac{d P(\tilde{R})}{d \tilde{R}} 
\nonumber \\
&&
{}+\frac{1}{3n\left(2n-1\right)} 
t_s^2 \kappa^2 \bar{\rho} 
\left[ \left( \alpha-1 \right) \left( 
\frac{1 \pm \sqrt{1-4\tilde{R}^{-1/2}}}{1 \mp \sqrt{1-4\tilde{R}^{-1/2}}} 
\right) \right]^{-3n(1+w)}\,,
\label{eq:3.9}
\end{eqnarray}
respectively, 
where $\tilde{t}_s \equiv t_s/\left[ 6n\left( 2n-1 \right) \right]$. 
Here, $\bar{\rho}$ corresponds to the 
present energy density of the matter. In particular, 
we use the present value of cold dark matter with $w=0$ for 
$\bar{\rho}$, i.e., $\bar{\rho} = 0.233 \rho_\mathrm{c}$~\cite{WMAP}, 
where $\rho_\mathrm{c} =3H_0^2/\left(8 \pi G \right) = 
3.97 \times 10^{-47} \mathrm{GeV}^4$ is the critical energy density 
and $H_0=2.13h \times 10^{-42}$GeV~\cite{Kolb and Turner} 
with $h =0.70$~\cite{Freedman:2000cf, Riess:2009pu} is the present Hubble 
parameter. From 
Eqs.~(\ref{eq:2.4}) and (\ref{eq:3.7}), we have 
\begin{eqnarray}
\frac{F(\tilde{R})}{2\kappa^2} = \frac{1}{2\kappa^2 \tilde{t}_s^{\,2}}
\left( P(\tilde{R}) \tilde{R} + \tilde{t}_s^{\,2} Q(\tilde{R}) \right)\,.
\label{eq:3.10}
\end{eqnarray} 
We examine $F(\tilde{R})$ by solving Eqs.~(\ref{eq:3.8})--(\ref{eq:3.10}) 
numerically. 

In Figs.~2 and 4, we depict $P(\tilde{R})$ and $\tilde{t}_s^{\,2} Q(\tilde{R})$ as functions of $\tilde{R}$. We have used $t_0 \approx 1/H_0$. 
Figs.~2 and 4 show the case of 
$0<t<t_\mathrm{c} (= t_0)$, i.e., $0<X<1/2$ and 
that of $t_\mathrm{c} (= t_0) < t<t_s$, i.e., $1/2<X<1$, respectively. 
We have numerically solved Eq.~(\ref{eq:3.8}) in the range of $\tilde{R}$ as 
$16.0001 \leq \tilde{R} \leq 18.0$. 
Here, we have taken the initial conditions as $P(\tilde{R} = 16.0001) =1.0$ 
and $d P(\tilde{R} = 16.0001)/ ( d \tilde{R} ) =0$ so that around the 
time when a crossing of the phantom divide occurs $t_\mathrm{c}$, 
$F(R)/\left( 2\kappa^2 \right)$ could contain the term 
$R/\left( 2\kappa^2 \right)$, namely, the ordinary Einstein-Hilbert action. 
We note that at $t= t_\mathrm{c}$, i.e., $X=1/2$ and hence $\tilde{R} = 16.0$, 
we cannot solve Eq.~(\ref{eq:3.8}) numerically. 
We therefore investigate the behavior of $F(R)$ gravity around a 
crossing of the phantom divide. 
By using Eq.~(\ref{eq:3.10}), 
we show the behavior of $F(\tilde{R})/\left(2 \kappa^2 \right)$ in Figs.~3 and 
5, which show the case of $0<t<t_\mathrm{c} (= t_0)$, i.e., $0<X<1/2$ and 
that of $t_\mathrm{c} (= t_0) < t<t_s$, i.e., $1/2<X<1$, respectively. 
For $\alpha = 2.0$, because $t_\mathrm{c} = t_0$, $0<X<1/2$ and 
$1/2<X<1$ correspond to the past and the future, respectively. 
Furthermore, we illustrate $w_\mathrm{eff} (\tilde{R}) = 
-1 +\left[2/\left(3n \right) \right] \left( 1-2X(\tilde{R}) \right)$ 
in Fig.~6. The time evolution of $\tilde{R}$ is given in Fig.~7. From 
Figs.~6 and 7, we see that a crossings of the phantom divide can 
be realized. 
The results in the all figures are shown by dimensionless 
quantities. 

\begin{figure}[tbp]
\begin{center}
   \includegraphics{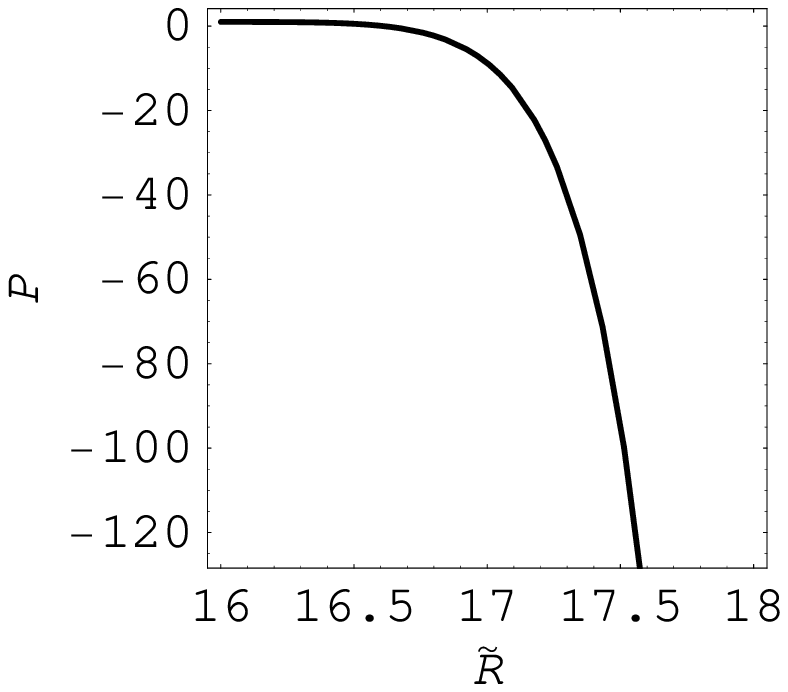}
   \includegraphics{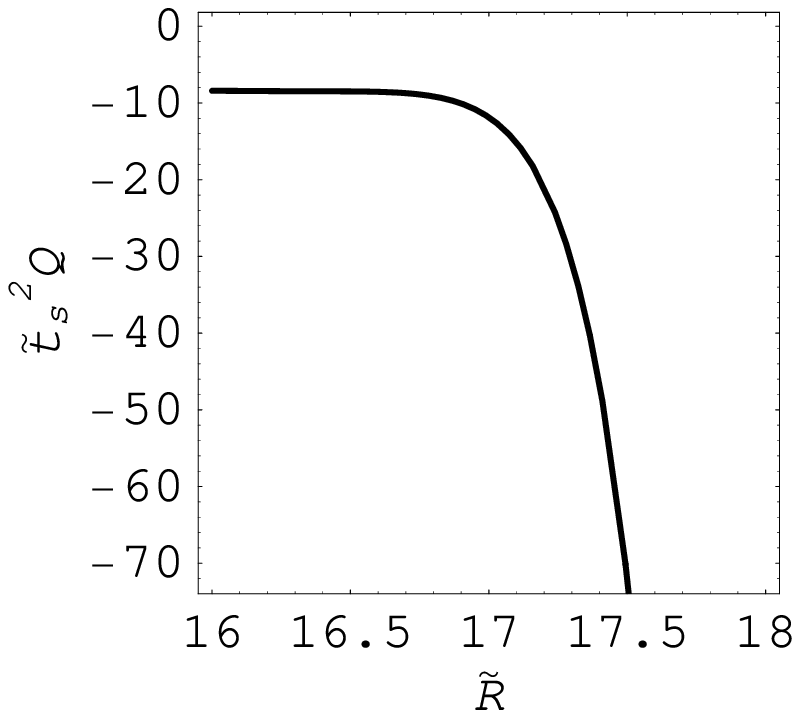}
\caption{$P(\tilde{R})$ and $\tilde{t}_s^{\,2} Q(\tilde{R})$ as 
functions of $\tilde{R}$ for $0<t<t_\mathrm{c} (= t_0)$. 
We have taken $n=10$, $\alpha = 2.0$ and $\bar{\rho} = 0.233 \rho_\mathrm{c}$. 
}
\end{center}
\label{fg:2}
\end{figure}

\begin{figure}[tbp]
\begin{center}
   \includegraphics{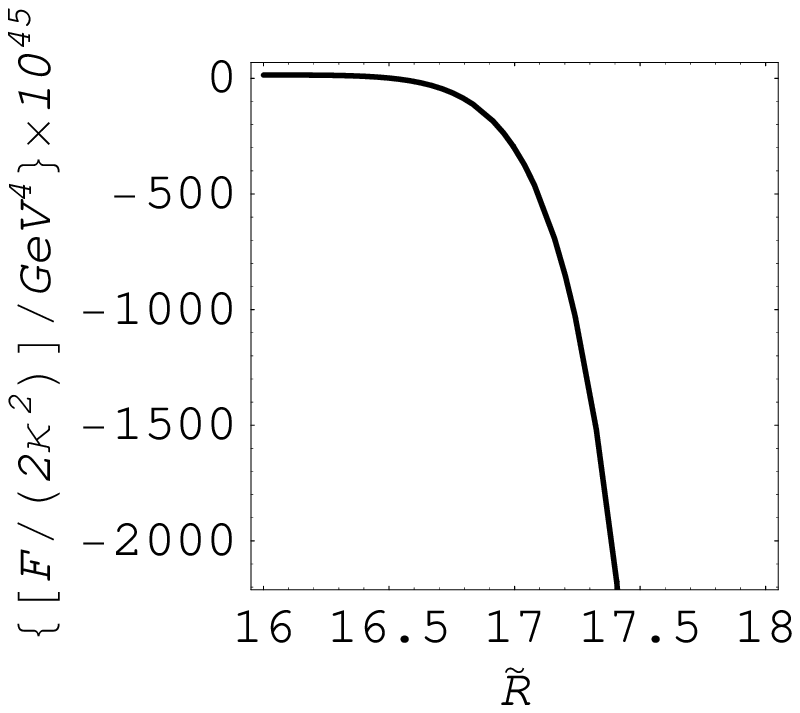}
\caption{
Behavior of $F(\tilde{R})/\left(2 \kappa^2 \right)$ as a function of 
$\tilde{R}$ for $0<t<t_\mathrm{c} (= t_0)$. 
Legend is the same as Fig.~2.
}
\end{center}
\label{fg:3}
\end{figure}

\begin{figure}[tbp]
\begin{center}
   \includegraphics{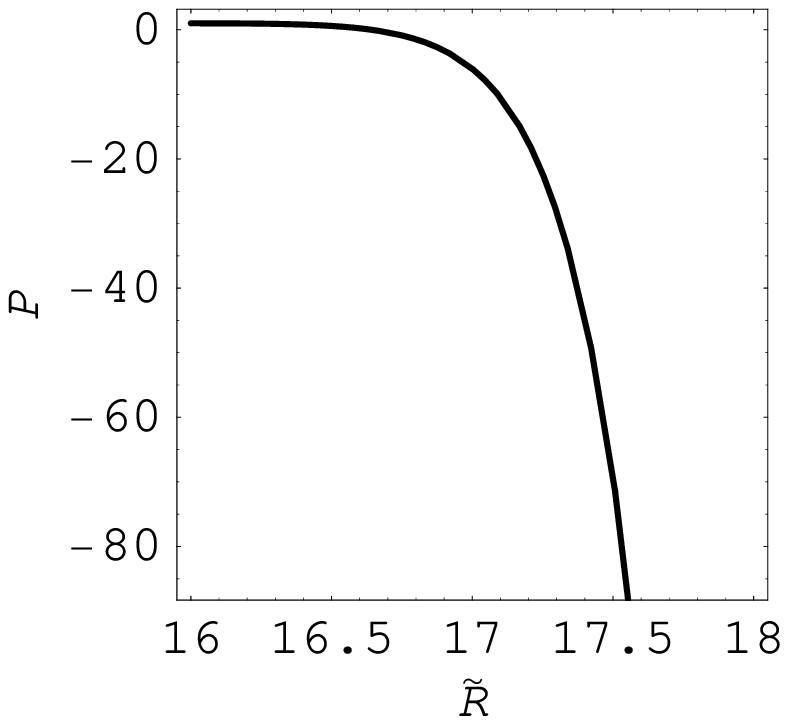}
   \includegraphics{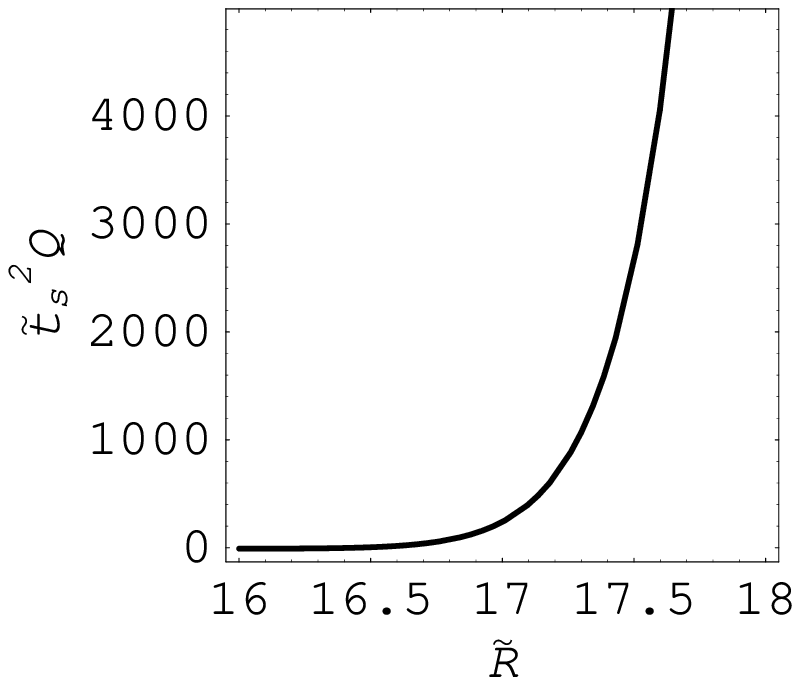}
\caption{$P(\tilde{R})$ and $\tilde{t}_s^{\,2} Q(\tilde{R})$ as 
functions of $\tilde{R}$ for $t_\mathrm{c} (= t_0) < t<t_s$.  
We have taken $n=10$, $\alpha = 2.0$ and $\bar{\rho} = 0.233 \rho_\mathrm{c}$. 
}
\end{center}
\label{fg:4}
\end{figure}

\begin{figure}[tbp]
\begin{center}
   \includegraphics{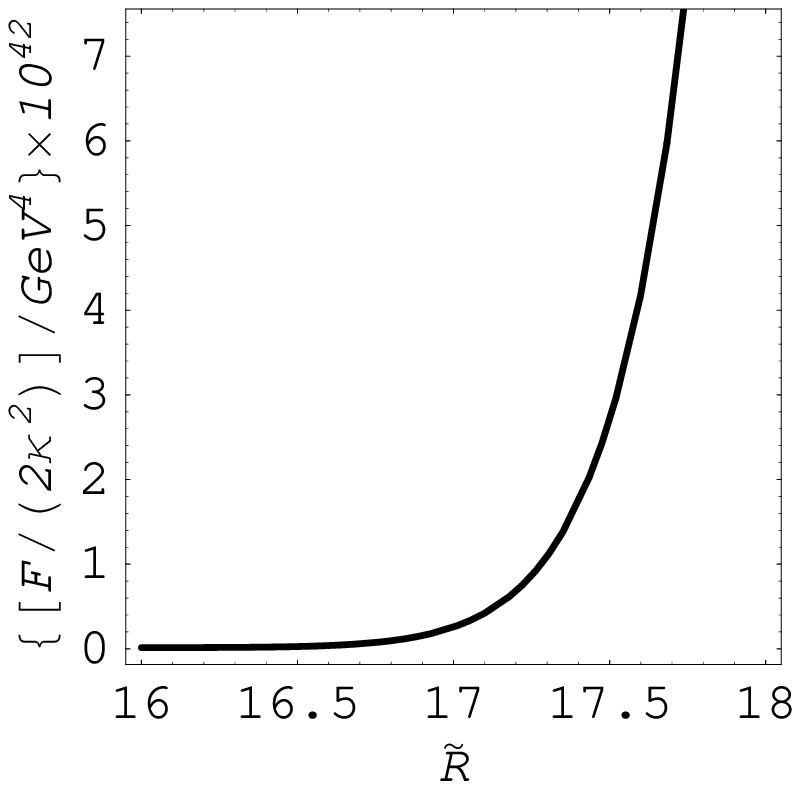}
\caption{
Behavior of $F(\tilde{R})/\left(2 \kappa^2 \right)$ as a function of 
$\tilde{R}$ for $t_\mathrm{c} (= t_0) < t<t_s$. 
Legend is the same as Fig.~4.
}
\end{center}
\label{fg:5}
\end{figure}

\begin{figure}[tbp]
\begin{center}
   \includegraphics{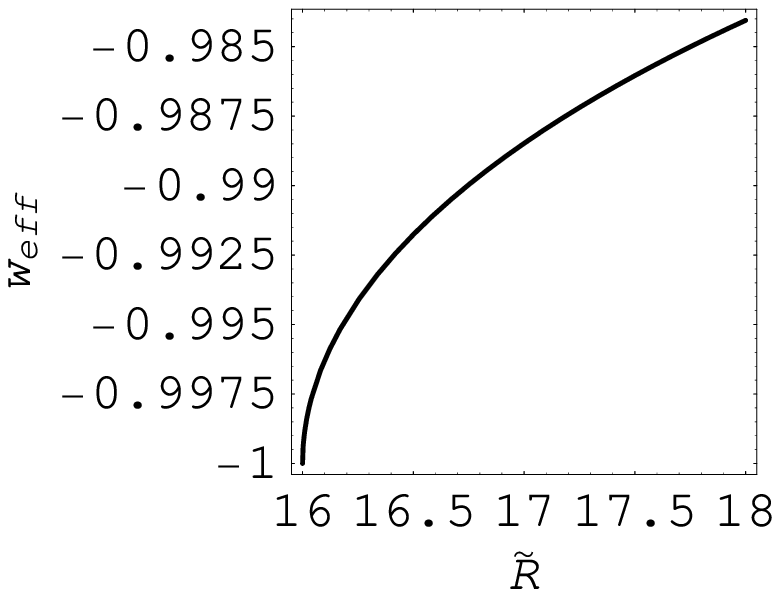}
   \includegraphics{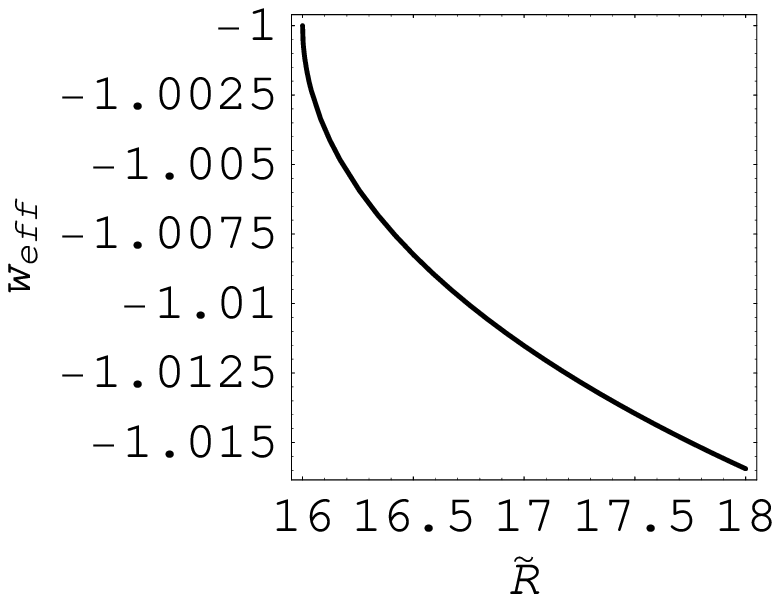}
\caption{
Behavior of $w_\mathrm{eff}(\tilde{R})$ as a function of 
$\tilde{R}$. 
The left panel and the right one show the case of 
$0<t \leq t_\mathrm{c} (= t_0)$, i.e., $0<X \leq 1/2$ and 
that of $t_\mathrm{c} (= t_0) \leq t<t_s$, i.e., $1/2 \leq X<1$, respectively. 
Legend is the same as Fig.~1.
}
\end{center}
\label{fg:6}
\end{figure}

\begin{figure}[tbp]
\begin{center}
   \includegraphics{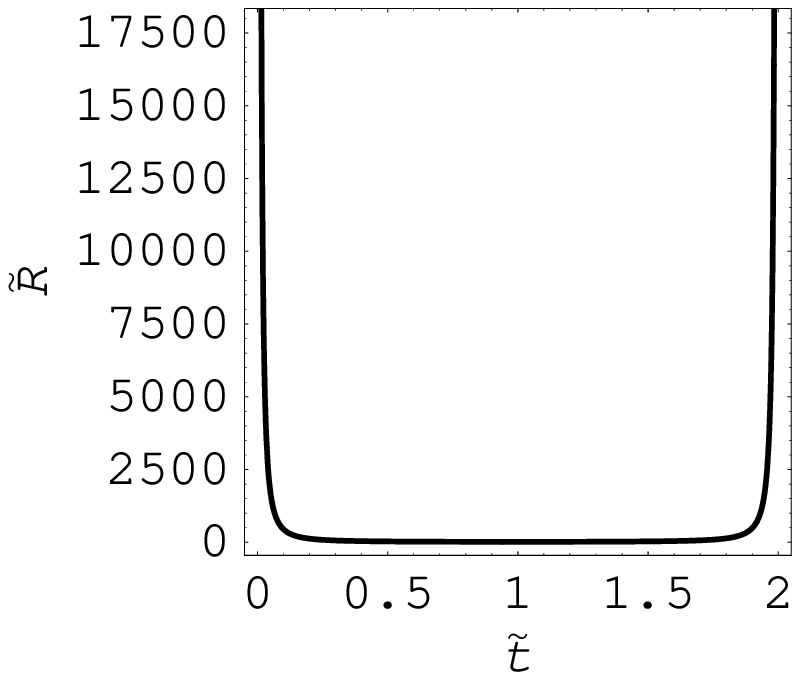}
\caption{
Time evolution of $\tilde{R}$. 
Legend is the same as Fig.~1.
}
\end{center}
\label{fg:7}
\end{figure}

 From Figs.~3 and 5, we see that before a 
crossing of the phantom divide, $F(\tilde{R})$ decreases in terms of 
$\tilde{R}$, while that after the crossing, $F(\tilde{R})$ increases in terms 
of $\tilde{R}$. 
The latter behavior is reasonable because 
in the Hu-Sawicki model~\cite{Hu:2007nk} of $F(R)$ gravity, which 
passes the solar system tests, $F(R)$ increases around the present 
curvature. 
For $\alpha = 2.0$, the time when a crossings of the phantom divide is 
the present time, and it follows from Eqs.~(\ref{eq:3.5}) and (\ref{eq:3.7}) 
that the present value of $\tilde{R}$ is $\tilde{R}_0 = 16.0$. 
We remark that such behavior is typical for a general class of viable 
$F(R)$ gravities introduced in Ref.~\cite{VNO} to which 
the Hu-Sawicki model belongs. This class of modified gravities can satisfy 
the solar system tests and unify inflation with the late-time cosmic 
acceleration. 
As viable models of $F(R)$ gravity, e.g., the models in 
Refs.~\cite{VNO, Starobinsky:2007hu, Appleby:2007vb, Amendola:2006we, 
Li:2007xn, Tsujikawa:2007xu, Capozziello:2007eu} are also proposed. 
In addition, investigations to solve the problem of a curvature singularity 
in $F(R)$ gravity~\cite{Abdalla:2004sw, Tsujikawa:2007xu, Briscese:2006xu, 
Appleby:2008tv, Frolov:2008uf, Bamba:2008ut, Kobayashi:2008tq} have recently 
been executed in Refs.~\cite{Nojiri:2008fk, Dev:2008rx, Kobayashi:2008wc, 
Capozziello:2009hc, Babichev:2009td, Upadhye:2009kt, 
Thongkool:2009js}. 
Theories without a singularity were also constructed 
in Ref.~\cite{Abdalla:2004sw}. 

Finally, we mention the stability for the obtained solutions of the phantom 
crossing under a quantum correction coming from the conformal anomaly. 
In Ref.~\cite{Bamba:2008hq}, it has been shown that 
the quantum correction of massless conformally-invariant fields could be small 
when a crossings of the phantom divide occurs and therefore the solutions of 
the phantom crossing could be stable under the quantum correction, 
although the quantum correction becomes important near the Big Rip singularity.

\section{Conclusion}

In the present paper, we have investigated a model of $F(R)$ gravity in which 
a crossing of the phantom divide can be realized. 
In particular, we have illustrated the behavior of $F(R)$ gravity around a 
crossing of the phantom divide by taking into account the presence of cold 
dark matter. 
The demonstration in this work can be interpreted as a meaningful 
step to consider a more realistic model of $F(R)$ gravity, which could 
correctly describe the expansion history of the universe.

\section*{Acknowledgments}

The author thanks Professor Chao-Qiang Geng, Professor Shin'ichi Nojiri and 
Professor S.~D.~Odintsov for their collaboration in 
Ref.~\cite{Bamba:2008hq} and important comments very much. 
He is also grateful to Professor Hideo Kodama for very helpful 
discussion of related problems. 
In addition, he acknowledges the KEK theory exchange program 
for physicists in Taiwan and the very kind hospitality of 
KEK. 
This work is supported in part by
the National Science Council of R.O.C. under:
Grant \#s: NSC-95-2112-M-007-059-MY3 and
National Tsing Hua University under Grant \#:
97N2309F1 (NTHU).


\end{document}